\documentstyle[epsfig,pennames]{camera}
\newcommand{\ipb}{\mbox{$\mathrm {pb}^{-1}$}}

\newcommand{\GEVcc}{\mbox{$\mathrm{Ge\!V}/{c^2}$}}
\newcommand{\Pepem}{\mbox{$\mathrm{e}^+\mathrm{e}^-$}}

\newcommand{\ev}{\mbox{$\mathrm{e}\nu$}}

\newcommand{\ee}{\mbox{$\mathrm{e}\mathrm{e}$}}
\newcommand{\PZ}{\mbox{${\rm Z}$}}
\newcommand{\MW}{\mbox{$m_{\mathrm{W}}$}}

\def\ra{\rightarrow} 
\def\rs{\mbox{$\sqrt{s}$}} 
\def\Journal#1#2#3#4{{#1} {\bf #2}, #3 (#4)}

\def\NPB{{\em Nucl. Phys.} B}
\def\PLB{{\em Phys. Lett.}  B}

\def\PRD{{\em Phys. Rev.} D}

\def\EPJC{{\em Eur. Phys. J.} C}
\def\CPC{\em Comp. Phys. Commun.}

\begin{document}

%
\title{FOUR-FERMION FINAL STATES \\ 
IN e$^+$e$^-$ COLLISIONS AT LEP2}

%
\author{Paolo Azzurri}

%
\organization{European Organization for Nuclear Research\\
CERN, EP Division, CH-1211 Gen\`eve 23, Switzerland.}

\maketitle

%
\abstract{The measurements of four-fermion production rates 
performed at LEP2 in $\Pepem$ collisions at centre-of-mass energies 
ranging from 161 to 209~GeV are presented. The focus is put 
on processes that involve the production and decay 
of W or Z electroweak bosons. 
Results on W decay rates and couplings are also discussed.}

\section{Introduction}
Between 1995 and 2000, the LEP collider at CERN 
delivered $\Pepem$ collisions above the Z peak, 
at centre-of-mass energies up to 209~GeV, 
corresponding to an integrated luminosity of about 700~\ipb
for each of the four experiments.
Within the Standard Model (SM) of electroweak interactions, the most
interesting new processes in this energy range 
are those leading to four-fermion final states, 
arising from single or pair productions 
W or Z bosons.
For each process the average
ratio (${\cal R}$) between the measured cross sections and the SM
expectations are determined, in order to illustrate their level of agreement.

\section{Single-W and single-Z productions}
For the single-boson production  
($\Pepem\ra\PW\ev,\PZ\Pe\Pe$), a subset of 
Feynman diagrams leading to identical four-fermion
final states, defines the signal to be measured.
The signal is further defined with additional kinematic cuts on the
outgoing four-fermion phase-space configuration~\cite{ewco}.  
The selections yield efficiencies of 40 to 60\%
with purities of 70 to 50\%, depending on the decay channel.
\begin{figure}[h]
  \begin{center}
\begin{picture}(100,60)(0,-25)
\put(15,25){(a)} \put(95,25){(b)}
\begin{minipage}[r]{4cm} 
\epsfig{file=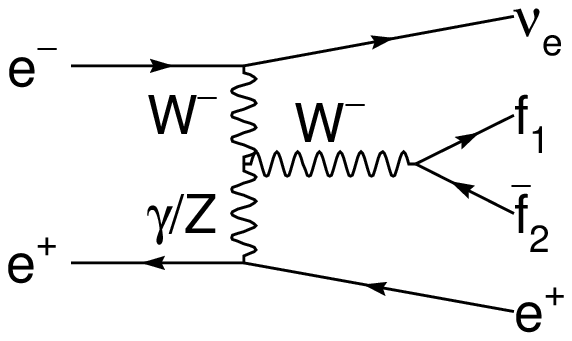,height=2.5cm}
\epsfig{file=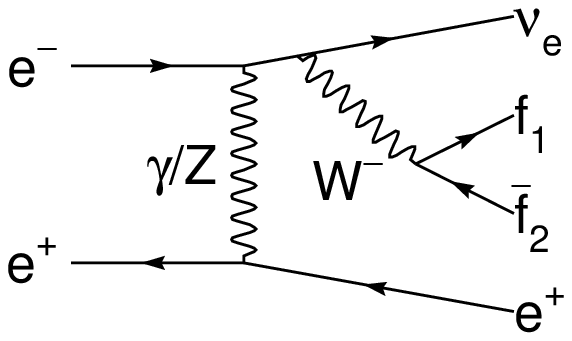,height=2.5cm}
\end{minipage}
\begin{minipage}[l]{6cm}
\epsfig{file=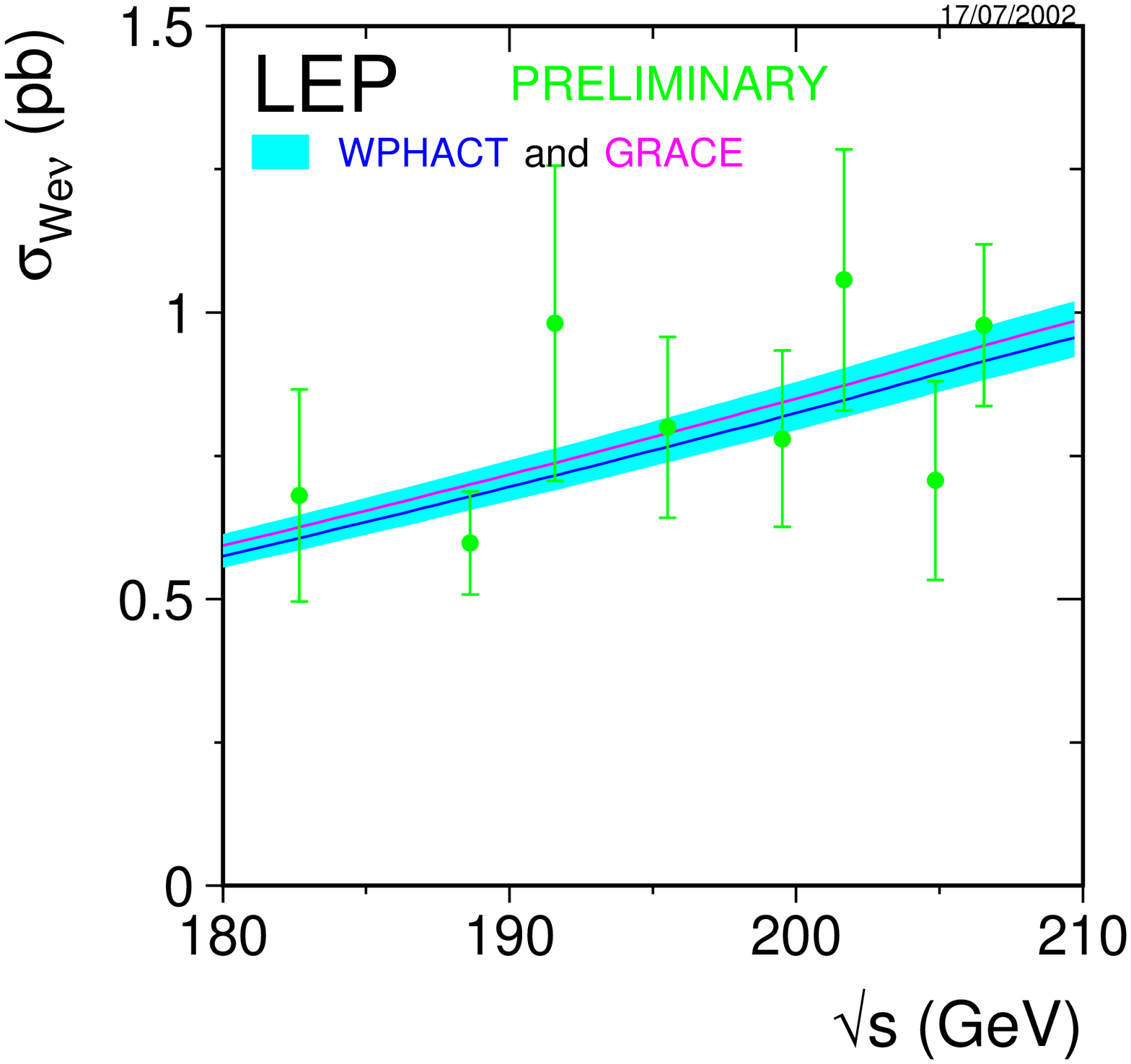,height=7cm}
\end{minipage}
\end{picture}
\end{center}
\caption{\footnotesize
Electroweak $t$-channel diagrams (a) and combined LEP cross section 
measurements at \rs=183-209~GeV (b), 
for the $\Pepem\ra\PW\ev$ process.}
\label{fig:wev}
\end{figure}
The measured cross sections 
for single-W~\cite{ewco} production (Fig.~\ref{fig:wev}b), 
and for single-Z~\cite{zee} productions are in good agreement 
with the SM predictions,
as calculated by the {\tt grace} program~\cite{grace},
within the accuracy of the measurements, of 8\% and 9\% respectively.
The resulting cross section ratios are 
${\cal R}_{\PW\ev}=0.949\pm 0.078$ and 
${\cal R}_{\PZ\ee}=0.928\pm 0.088$.
The single-W cross section measurement also allows the WW$\gamma$
trilinear gauge-boson coupling to be constrained with an accuracy 
of 15\%~\cite{ewco}.

\begin{figure}[h]
\centerline{
\begin{picture}(100,65)(0,-30)
\put(15,27){(a)} \put(98,27){(b)}
\begin{minipage}[r]{4cm}
\epsfig{file=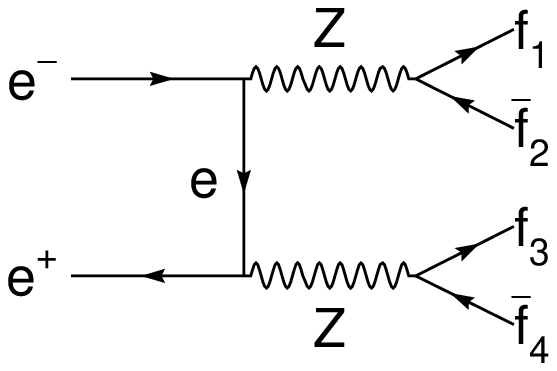,height=2.5cm}
\epsfig{file=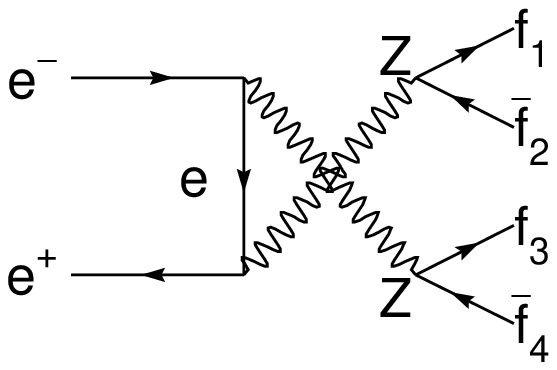,height=2.5cm}
\end{minipage}
\begin{minipage}[r]{6cm}
\epsfig{file=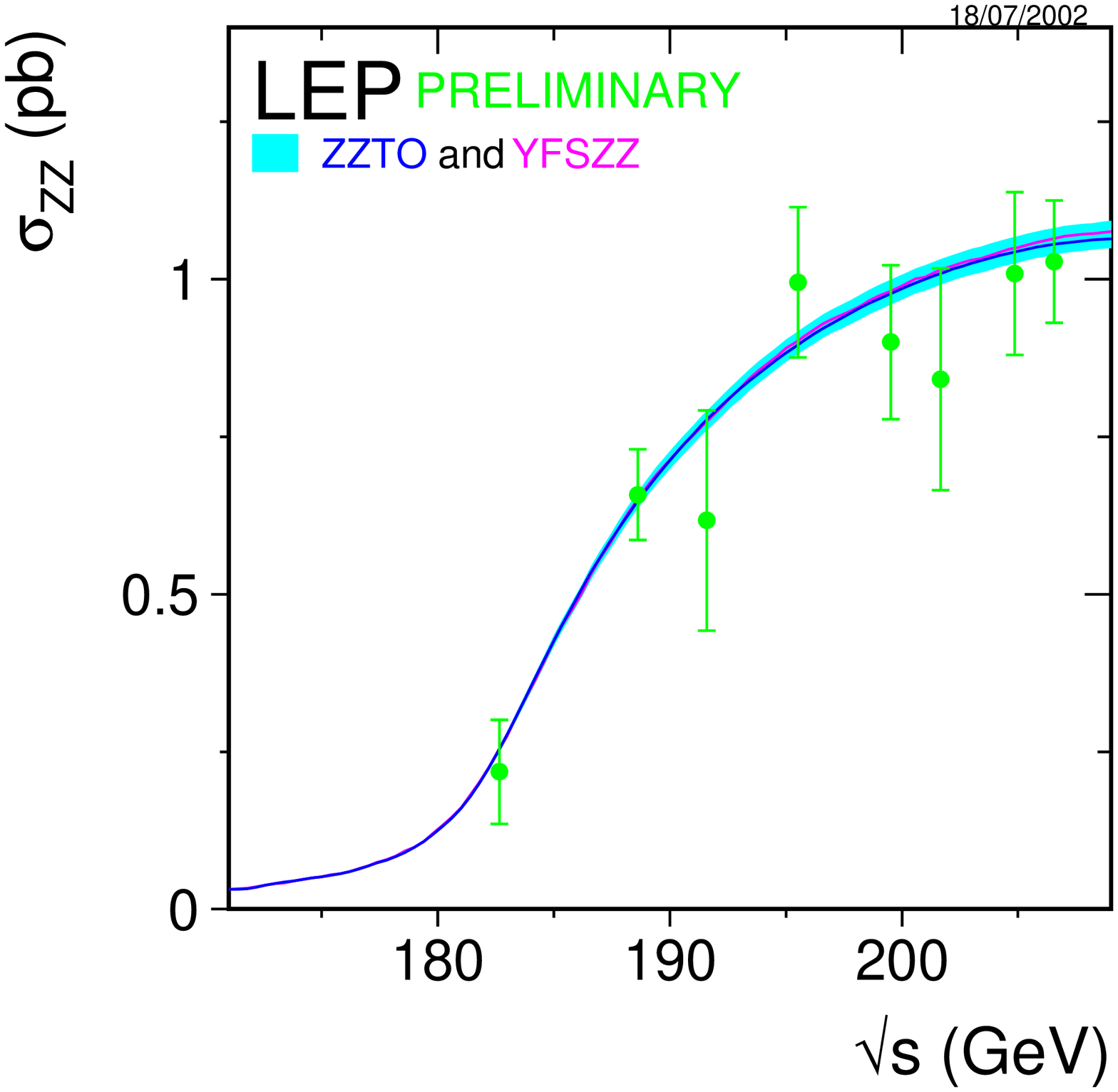,height=7.5cm}
\end{minipage}
\end{picture}
}
\caption{\footnotesize
Electroweak {\tt NC02} diagrams (a) and combined LEP cross section 
measurements at \rs=183-209~GeV (b), 
for the $\Pepem\ra\PZ\PZ$ process.}
\label{fig:zz}
\end{figure}
\section{Z-pair and Z$\gamma^\ast$ productions}

The Z-pair production measurements are defined 
as the contribution of the 
{\tt NC02} diagrams (Fig.\ref{fig:zz}a)
to the detected four-fermion final states. 
Selection efficiencies of 50 to 80\% are achieved
with purities of 80 to 50\%, depending on the decay channel.
The measured Z-pair cross sections, 
(Fig.\ref{fig:zz}b),
are in agreement with the theoretical predictions
of {\tt yfszz}~\cite{yfszz}, 
within the measurements accuracy of 5.5\%.
The resulting cross section ratio is 
${\cal R}_{\PZ\PZ}=0.962\pm 0.055$.
Rates for Z$\gamma^\ast$ four-fermion processes are also measured 
to be in agreement with the SM predictions~\cite{zgs}.

\section{W-pair productions}
\begin{figure}[h]
\centerline{
\begin{picture}(100,70)(0,-30)
\put(15,27){(a)} \put(98,27){(b)}
\begin{minipage}[r]{4cm}
\epsfig{file=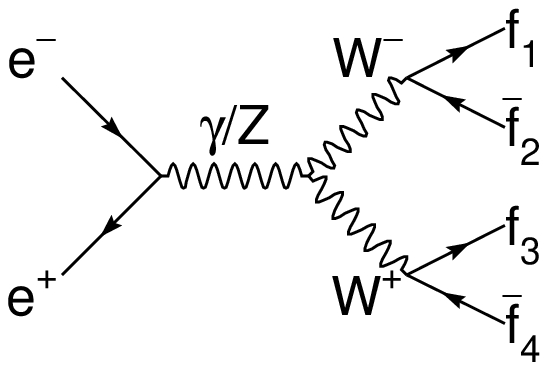,height=2.5cm}
\epsfig{file=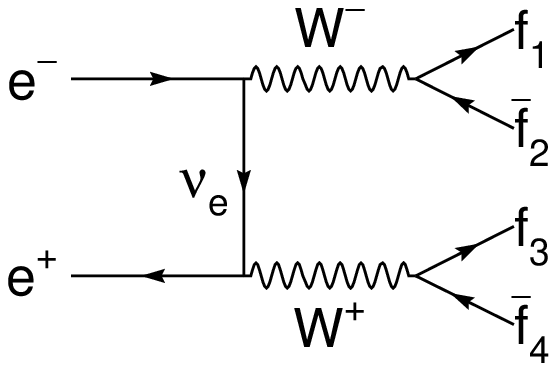,height=2.5cm}
\end{minipage}
\begin{minipage}[r]{6cm}
\epsfig{file=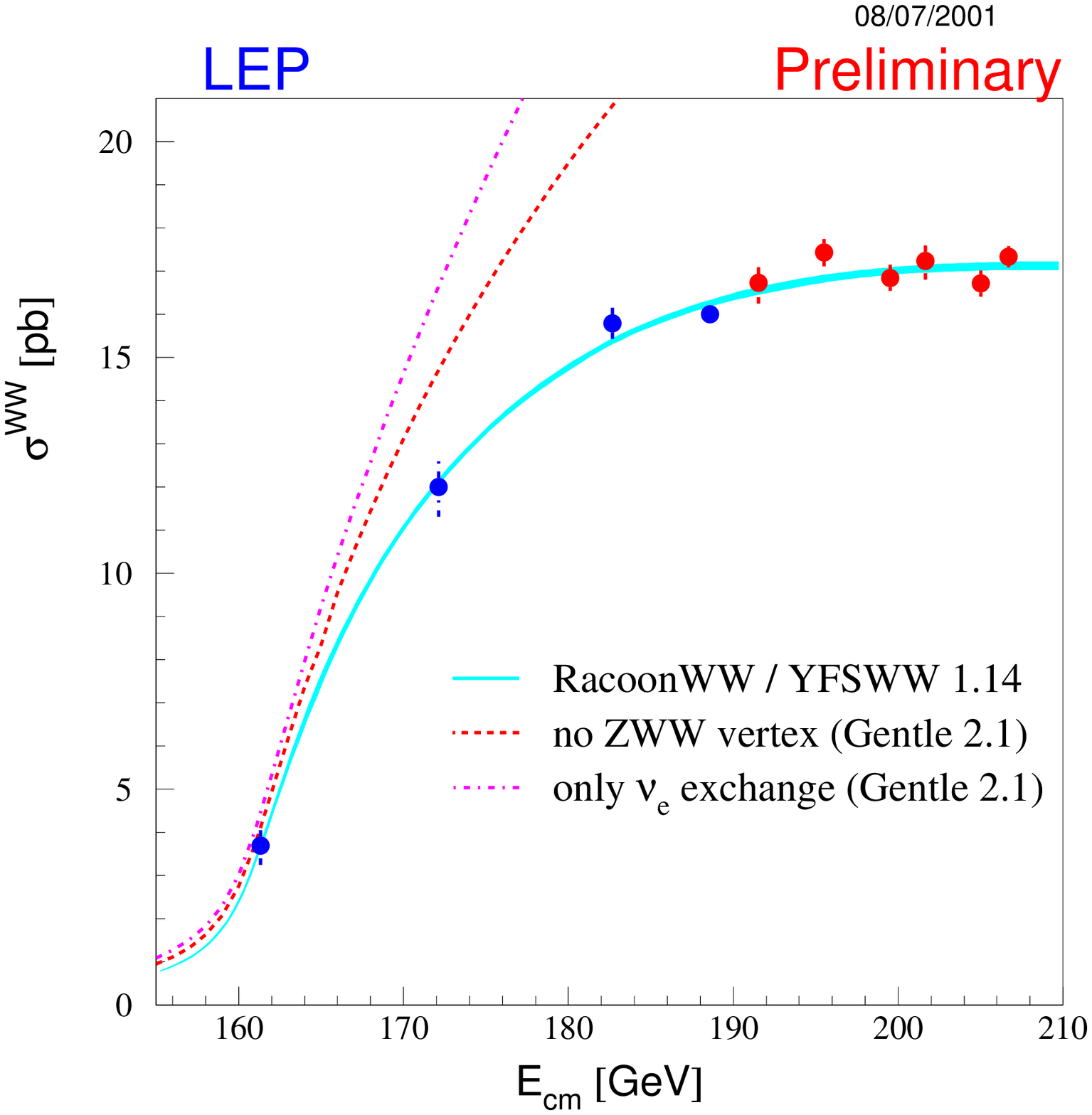,height=7.5cm}
\end{minipage}
\end{picture}
}
\caption{\footnotesize
Electroweak {\tt CC03} diagrams (a) and LEP combined cross section 
measurements at \rs=161-209~GeV (b), for the $\Pepem\ra\PWp\PWm$ process.
The experimental points clearly prove the existence of both the 
WW$\gamma$ and WWZ SU(2)$\otimes$U(1) gauge-boson self-couplings.}
\label{fig:ww}
\end{figure}

The LEP W-pair production is defined as the contribution
of the {\tt CC03} diagrams (Fig.\ref{fig:ww}a)
to the detected four-fermion final states. 
Selection efficiencies vary from 60 to 90\%
with purities of 90 to 80\%, depending on the decay channel.
The W-pair cross section experimental determinations
at \rs=161-209~GeV (Fig.\ref{fig:ww}b), are in agreement with 
the theoretical predictions of {\tt yfsww}~\cite{yfsww}
and {\tt racoonww}~\cite{racoonww} to an
overall precision of 1.1\%. 
This precision allowed the first clear proof 
of the existence of WW$\gamma$ and WWZ couplings to be obtained,
as visible in Fig.~\ref{fig:ww}b, 
and to test the effects of O($\alpha$) electroweak radiative 
corrections to the {\tt CC03} diagrams. 
The resulting cross section ratio is 
${\cal R}_{\PW\PW}=0.997\pm 0.011$.

The measurement of the W-pair cross section at the production
threshold, is sensitive to the W mass value,
and allowed the determination of 
$\MW= 80.40 ^{+0.22}_{-0.21}~\GEVcc$ to be achieved,
independently from the subsequent direct mass determinations
performed at LEP2.

\section{W decay rates and couplings}
The sample of W decays collected at LEP 
(about 10$^4$ WW events per experiment), allowed for the first time
the hadronic and leptonic W decay branching ratios to be 
determined directly.
The combined value BR(\PW$\ra$hadrons)=67.92$\pm$0.27\%
is in agreement with the SM expectations, and 
tests the universality of lepton and quark charged weak couplings
at the 0.7\% level, $g_q/g_\ell= 1.010\pm 0.007$.
The W hadronic branching ratio measurement is also used to 
constrain the CKM quark-mixing matrix, its unitarity in the 
first two families, and to extract the best current constraint on the  
$V_{\mathrm cs}$ amplitude, $|V_{\mathrm cs}| = 0.996\pm0.013$.
Finally, the agreement of the measurement of the 
three leptonic W branching ratios
translates to a test of the universality of the charged weak couplings
of the three lepton species at the 1\% level, yielding
$g_\mu/g_{\rm e}  = 1.000\pm 0.010$,
$g_\tau/g_{\rm e}  =  1.026\pm 0.014$ and 
$g_\tau/g_{\mu}  =   1.026\pm 0.014$.

\section{Conclusions}
The measurement of four-fermion production rates at LEP
has been a nice and exciting work. It has provided different 
and crucial new results  
on the structure of gauge boson couplings, the W mass,
the lepton and quark universality of charged weak
interactions, and quark mixing.

\section*{Acknowledgements}
I would like to thank the organisers of the Parma XIV - IFAE 
meeting for providing a nice atmosphere and a fine organization.
I also thank G.~Sguazzoni for his pleasant company. 
Finally, I have no alternative but to thank P.~Janot 
for his critical reading of this manuscript.

%

\begin{thebibliography}{9}
\bibitem{ewco} The LEP and SLD Collaborations, {\sl A Combination of 
Preliminary Electroweak Measurements and Constraints on the Standard Model}, 
LEPEWWG/2002-01; CERN-EP/2001-098 (Dec.2001), and references therein;
({\sf http://lepewwg.web.cern.ch/LEPEWWG/}).
\bibitem{zee} 
OPAL Collaboration, CERN-EP/2002-052; \Journal{\EPJC}{24}{1}{2002}. 
DELPHI Collaboration, conference note 591 (2002-057);
\Journal{\PLB}{515}{238}{2001}. 
L3 Collaboration, conference note 2771 (2002).
ALEPH Collaboration, conference note 2002-018.
\bibitem{grace} Minami-Tateya Collaboration, 
  J. Fujimoto {\it et al.}, \Journal{\CPC}{100}{128}{1997};
  F. Yuasa {\it et al.}, 
  \Journal{\em Prog. Theor. Phys. Suppl.}{138}{18}{2000}; hep-ph/0007053.
  {\sf http://www-sc.kek.jp/minami/}.
\bibitem{yfszz}
S. Jadach, W. P\l aczek, M. Skrzypek, B.F.L. Ward and Z. W\c as,
 \Journal{\PRD}{56}{6939}{1997}
\bibitem{zgs} DELPHI Collaboration, conference note 524 (2001-096).
\bibitem{yfsww} 
S. Jadach, W. P\l aczek, M. Skrzypek, B.F.L. Ward and Z. W\c as,
 \Journal{\PRD}{54}{5434}{1996}; \Journal{\PLB}{417}{326}{1998}; 
 \Journal{\PRD}{61}{113010}{2000}.
\bibitem{racoonww} A. Denner, S. Dittmaier, M. Roth and D. Wackeroth,
 \Journal{\NPB}{560}{33}{1999}; \Journal{\NPB}{587}{67}{2000};
 \Journal{\PLB}{475}{127}{2000}; hep-ph/0101257 (2001).

\end{thebibliography}
\end{document}